\documentstyle [prl,twocolumn,aps] {revtex}
\begin{document}
\draft
\title{ Kinetic model for q-deformed bosons and fermions }
\author{ G. Kaniadakis, A. Lavagno and P. Quarati}
\address{ Dipartimento di Fisica and INFM- 
Politecnico di Torino \\
Corso Duca degli Abruzzi 24, 10129 Torino, Italy \\ 
Istituto Nazionale di Fisica Nucleare, Sezioni di 
Cagliari e di Torino}
\maketitle
\begin {abstract} {\bf Abstract:} 
We have studied the kinetics of 
$q$-deformed bosons and fermions, within a semiclassical approach. 
This investigation is realized by introducing a generalized exclusion-inclusion 
principle, 
intrinsically connected with the quantum $q$-algebra by means of the creation 
and annihilation operators matrix elements.
In this framework, we have derived a 
non-linear Fokker-Planck equation for $q$-deformed bosons and fermions 
which can be seen as a time evolution equation, appropriate to consider 
non-equilibrium or near-equilibrium systems in a semiclassical approximation. 
The steady state of this equation reproduces in a simple mode the 
$q$-oscillators equilibrium statistics.
\end {abstract}

Recently, there has been a great deal of interest in the study of the quantum 
groups and of their applications to $q$-deformed bosons and fermions. 
Several models \cite{bie,mac,ng} of $q$-deformed free bosons and fermions 
gas \cite{gre,lee,su,tus,wer,song,dao} have been applied to Bose gas 
condensation \cite{hsu,mont1}, anyon fields \cite{gol}, 
phonon spectrum in $^4 He$ \cite{mont2}, 
asymmetric XXZ Heisenberg chain \cite{alca}, nuclear and molecular physics 
\cite{ray,iswao,sha1,sha2}, conformal field theories \cite{alva} and 
quantum Lie superalgebras \cite{cha}.

The $q$-boson and $q$-fermion particles are defined by an appropriate 
deformation of the 
commutation-anticommutation relations of the creation and annihilation 
operators, modifying the exchange factor between permuted particles.

For $q$-boson particles, we have the $q$-deformed commutations \cite{ng,lee}

\begin{equation}
[a,a]=[a^{\dag},a^{\dag}]=0, \ \ \ aa^{\dag} - q a^{\dag} a =q^{-N} \ \ ;
\end{equation}
the Hilbert space with basis $\mid n>$ is constructed with these features
\begin{eqnarray}
&&a \mid 0>=0, \ \ \ a^{\dag} \mid n>=[n+1]^{1/2} \mid n+1> \ \ , \nonumber \\
&&a \mid n >=[n]^{1/2} \mid n-1>, \ \ \ N \mid n>=n \mid n> \ \ ,
\end{eqnarray}
implying the following operatorial relations
\begin{equation}
a^{\dag} a = [N], \ \ \ a a^{\dag}=[1+N] \ \ .
\end{equation}
The notation [ ] is defined in terms of the deformation parameter $q$ as 
\begin{equation}
[x]=\frac{q^x - q^{-x}}{q-q^{-1}} \ \ .
\end{equation}
In the limit $q\rightarrow 1$, $[x]\rightarrow x$ and one reproduces 
the familiar boson algebra.

For $q$-fermion operators $b$ and $b^{\dag}$, analogously to the boson case, 
we have the anticommutation relations \cite{ng,lee}
\begin{equation}
\{b,b\}=\{b^{\dag},b^{\dag}\}=0, \ \ \ bb^{\dag} + q^{-1} b^{\dag} b =
q^{-N} \ \ .
\end{equation}
The number operator $N$, for this case, can take only the values $n=0,1$ 
and the above anticommutations imply the following relations

\begin{eqnarray}
b \mid 0>= b^{\dag}\mid 1>=0, \ \ \  \mid 1>=b^{\dag} \mid 0>  
\end{eqnarray}
and, as a consequence, we have
\begin{equation}
b^{\dag} b = [N], \ \ \  b b^{\dag} = [1-N] \ \ .
\end{equation}

In the recent past several authors \cite{lee,tus,hsu,vok}, 
assuming the relations of Eqs.(3) and (7), have derived the 
statistical distributions of $q$-oscillators (let us recall at this point 
that the 
statistical distributions can not be derived from 
the thermal averages $<a^{\dag} a>$ or $<b^{\dag} b>$ because these 
quantities are the average numbers of particles only in the undeformed  
($q=1$) case).

In this letter we want to show that the thermodynamic and statistical 
properties of 
$q$-deformed bosons and fermions can be derived in the framework 
of a kinetic approach, recently proposed by us \cite{ka1,ka2}. 
The crucial point of our model is the introduction in the transition 
probability of an exclusion-inclusion Pauli principle which can be seen as the 
semiclassical consequence of the partial antisymmetrization-symmetrization 
of the quantum wave function. 
Assuming that the diffusional current must satisfy general conditions and 
that the exclusion-inclusion principle be valid 
in analogy to the pure fermion-boson case, we deduce a 
generalized non-linear Fokker-Planck equation appropriate to study the  
equilibrium and also the non-equilibrium conditions of $q$-oscillators.

Starting point of our approach is the Pauli master equation for the 
mean occupation function $n(t,v)$ in terms of the transition probability 
$\pi (t,v\rightarrow u)$ from the state $v$ to the state $u$. 
In the case of the one dimensional 
velocity space and in the limit of nearest-neighbor interaction, with 
infinitesimal transition from the state $v$ to the state $v+dv$, the master 
equation is
\begin{eqnarray}
\frac{\partial n(t,v)}{\partial t}  = &\pi&(t,v-dv\rightarrow v)+
\pi(t,v+dv \rightarrow v) \ \ \nonumber \\
-&\pi&(t,v \rightarrow v-dv)-\pi(t,v \rightarrow v+dv)\ \ .
\end{eqnarray}
The present formalism can be extended to a general dimension velocity space,  
for this generalization and a more exhaustive 
discussion of the above master equation we send to Ref.\cite{ka1}.

To introduce, from a semiclassical point of view, the quantum behavior 
of a generic $q$-oscillator boson or fermion system, 
we postulate a transition probability with a 
generalized exclusion-inclusion principle as \cite{ka1,ka2}

\begin{equation}
\pi(t,{ v} \rightarrow { u})=r(t,{ v},{ v}-{ u}) \, 
\varphi[n(t,{ v})] \,  \psi[n(t,{ u})] \ \ ,
\end{equation}
where $r(t,{ v},{ v}-{ u})$ is the transition rate from the state $v$ to the 
state $u$, $\varphi(n)$ and 
$\psi(n)$ are real functions that inhibit or enhance the transition 
probability from a site to another one. In fact, $\varphi(n)$ is a 
function depending on the occupational distribution at the initial state $v$ 
which must satisfy the condition $\varphi(0)=0$ (if the initial state is 
empty, the 
transition probability is equal to zero) and $\psi(n)$ is a function depending 
on the arrival state and satisfies the condition $\psi(0)=1$ (if the 
arrival state is empty, the transition probability is not modified). 
If we choose $\varphi(n)=n$ and $\psi(n)=1$ we find the standard 
classical linear kinetics.\\
Explicit expressions of the functions $\varphi(n)$ and $\psi(n)$ can be used to 
simulate  semiclassically a quantum exclusion-inclusion principle in the 
transition 
probability. In Ref. \cite{ka2,ka3} we have studied, in this framework, 
the generalized Haldane exclusion statistics \cite{hal} 
and the quantum extension of the 
non-extensive Tsallis entropy \cite{tsa}.

Because of the quantum meaning of the functions $\varphi(n)$ and $\psi(n)$, 
it must exist an intrinsic relation between these functions and the quantum 
algebra of the creation and annihilation operators matrix elements. 
As $\varphi(n)$ is proportional to the probability of finding in the state $v$ 
the occupational number $n$ and $\psi(n)$ is proportional to the probability of 
introducing an extraparticle into a state with occupational number $n$, we can 
postulate the following relations 

\begin{equation}
\varphi (n) \propto \mid < n -1 \mid a_n \mid n >\mid ^2 \ \ ,
\end{equation}

\begin{equation}
\psi (n) \propto \mid < n +1 \mid a_n^{\dag} \mid n >\mid ^2 \ \ ,
\end{equation}
implying an important connection between the semiclassical 
kinetic approach and the quantum algebra.

If we expand up to the second order in power of $dv$ the r.h.s. of Eq.(8), 
we obtain the non-linear generalized Fokker-Planck equation \cite{ka1,ka2}
\begin{eqnarray}
\frac{\partial n}{\partial t}=\frac{\partial}{\partial v} \Bigg [
\left (J+\frac{\partial D}{\partial v}\right ) \varphi(n) \psi(n) 
\ \ \ \nonumber \\
+ D \left (\psi(n) \frac{\partial \varphi(n)}{\partial n}-\varphi(n) 
\frac{\partial 
\psi(n)}{\partial n}\right ) \frac{\partial n}{\partial v} \Bigg ] \ \ ,
\end{eqnarray}
where the drift $J\equiv J(t,v)$ and the diffusion $D\equiv D(t,v)$ 
are respectively the first and the second order momenta of the transition rate.

Eq.(12) is a continuity equation and the factor into the square 
bracket is the particle current that can be seen as the sum of 
two terms. The first is the drift current $j_{_{drift}}$, given by 
\begin{equation}
j_{_{drift}}(t,v)=
\left (J+\frac{\partial D}{\partial v}\right ) \varphi(n) \psi(n) \ \ .
\end{equation}
This quantity is proportional to $\varphi(n) \psi(n)$, hence proportional 
to the probability $\pi(n)$ that the particle get away from the $v$ site. 
This is an anomalous current because it is a non-linear algebraic function 
of $n$. 
Only in the case of the standard Fokker-Planck kinetics 
($\varphi(n)=n$ and $\psi(n)=1$) this current is proportional to $n$. \\
The second term in Eq.(12) is a diffusion like current, $j_{_{diff}}$, 
proportional to $\partial n/\partial v$ and given by 
\begin{equation}
j_{_{diff}}(t,v)=D \, A(n) \, \frac{\partial n}{\partial v} =
{\cal D}(t,v,n) \, \frac{\partial n}{\partial v} \ \ , 
\end{equation}
with 
\begin{equation}
A(n)=\psi(n) \frac{\partial \varphi(n)}{\partial n}-\varphi(n) 
\frac{\partial \psi(n)}{\partial n} \ \ .
\end{equation}
We call $j_{_{diff}}$ anomalous diffusional current because the diffusional 
coefficient ${\cal D}(t,v,n)$ depends explicitly on $v$ and $n(t,v)$, 
while the standard diffusional coefficient does not depend on $n(t,v)$. 

Let us now limit ourselves to those physical processes described by Eq.(12)
where the diffusional current $j_{_{diff}}$ is a standard current or Fick 
current 
with the diffusional coefficient independent of the distribution function 
$n(t,v)$. 
To realize this condition it is necessary to determine the functions 
$\varphi(n)$ and $\psi(n)$ that satisfy the equation $A(n)=A$. 

In the case of a pure bosonic kinetics \cite{ka1}, 
one has $\varphi(n)=n$ and $\psi(n)=1+n$ or $\psi(n)=\varphi(\bar{n})$ with 
$\bar{n}=1+n$. 
For a pure fermionic kinetics \cite{ka1} 
one has $\varphi(n)=n$, $\psi(n)=1-n$  
and, in analogy with the bosonic notations, 
the fermionic condition can be written as $\psi(n)=\varphi(\bar{n})$, where now 
$\bar{n}=1-n$. \\
The condition $\psi(n)=\varphi(\bar{n})$ with $\bar{n}=1+\sigma n$ holds both 
for the pure bosonic kinetics ($\sigma=1$) and for 
the pure fermionic kinetics ($\sigma=-1$). 

Now, we impose that the above conditions are conserved also in the case 
the kinetics be defined by an arbitrary $\varphi(n)$ so that 
we limit ourselves to consider the physical processes which satisfy 
the two following general conditions 
\begin{eqnarray}
&&A(n)=A  \ \ , \nonumber \\
&&\psi(n)=\varphi(1+\sigma n) \ \ .
\end{eqnarray}
After derivation respect to $n$ of the first condition 
and taking into account the second condition in Eqs.(16), we easily obtain the 
following differential equation for the function $\varphi(n)$
\begin{equation}
\varphi(1+\sigma n) \varphi^{''}(n) 
- \varphi(n) \varphi^{''}(1+\sigma n)=0 \ \ .
\end{equation}
Introducing the auxiliary function $\lambda(n)=\varphi^{''}(n)/\varphi(n)$, 
it is easy to verify, by taking into account Eq.(17), 
that this function is equal to a constant $\alpha^2 \in {\rm I \! R}$ and our 
problem is reduced to the following Cauchy problem
\begin{eqnarray}
&&\varphi^{''}(n)-\alpha^2 \varphi(n)=0 \ \ , \nonumber \\
&&\varphi(0)=0 \ \ , \\
&&\varphi(1)=1 \ \ .\nonumber 
\end{eqnarray}
For $\alpha$ real the solution of Eq.(18) is given by 
\begin{equation}
\varphi(n)=\frac{\sinh\alpha n}{\sinh\alpha} \ \ ,
\end{equation}
while for $\alpha$ immaginary: $\alpha=i\vartheta$
\begin{equation}
\varphi(n)=\frac{\sin\vartheta n}{\sin\vartheta} \ \ .
\end{equation}
Eqs. (19) and (20) can be written using the definition 
of the symbol [ ] of Eq.(4) as 
\begin{equation}
\varphi(n)=[n] \ \ ,
\end{equation}
where $q=e^{\alpha}$ (for $q$ real we have the expression (19) and 
for $q$ complex we have Eq.(20)).\\
Thus, selecting $\psi(n)=\varphi(1+\sigma n)$ and imposing diffusional 
Fick current, we find the function $\varphi(n)$ and $\psi(n)$ for $q$-particles 
($q$-boson for $\sigma=1$ and $q$-fermion for $\sigma=-1$). 
These functions ($\varphi(n)=[n]$ and $\psi(n)=[1+\sigma n]$) correspond to the 
Eqs. (10) and (11), previously postulated, with the quantum $q$-algebra of 
Eqs.(1)-(7).

Now, if we include the $q$-functions $\varphi(n)$ and $\psi(n)$ in Eq.(13), 
in the case of Brownian particles ($J=\gamma v$, $D=\gamma/\beta m$, 
$\beta=1/kT$ and $\gamma$ is a dimensional constant), we obtain the 
Fokker-Planck equation for $q$-oscillators
\begin{eqnarray}
\frac{\partial n}{\partial t}=\frac{\partial}{\partial v} \Bigg [
\gamma v \, [n] \, [1+\sigma n] 
+ \gamma \frac{k T}{m} \frac{\alpha}{\sinh\alpha} 
\frac{\partial n}{\partial v} \Bigg ] \ \ .
\end{eqnarray}
This equation describes the time evolution of $q$-deformed bosons and fermions 
in the velocity space. 

Great interest is actually 
devoted to study the statistical distribution and the dynamical evolution 
toward equilibrium of several physical systems. 
However, because of the huge complexity of the 
kinetic transport theory on a quantum field theoretical basis, it results very 
difficult to describe the non-equilibrium behavior 
of physically interesting systems. \\
Because the quantum dynamics is equivalent to classical dynamics with the 
inclusion of quantum fluctuations, that in average coincide with the Brownian 
fluctuations, we conclude that the non-linear kinetic equation (22) can 
describe very close the quantum dynamics of a system of $q$-oscillators. 
Consequently, the above kinetic approach can result appropriate to 
describe non-equilibrium or near-equilibrium $q$-deformed systems with small 
(average) quantum fluctuations. These phenomena are actually very important in 
several applications from condensed matter to cosmological and high 
energy physics \cite{alva2,kelly,brandt}.

In stationary conditions ($t\rightarrow\infty$), the particle current in the 
square bracket vanishes and Eq.(22) becomes a homogeneous first-order 
differential equation, easily integrated as 
\begin{equation}
\frac{[n]}{[1+\sigma n]}=e^{-\epsilon} \ \ ,
\end{equation}
where $\epsilon=\beta (E-\mu)$, $E=\frac{1}{2} m v^2$ is the kinetic energy and 
$\mu$ is the chemical potential which can be evaluated by fixing the number 
of particles of the system.\\
Eq.(23) defines implicitly the statistical distribution $n(v)$ in stationary 
conditions; it can be explicitly derived as 
\begin{equation}
n=\frac{1}{\log q} \tanh^{-1} \left ( \frac{\sinh\log q}{e^{\epsilon}- 
\sigma \cosh\log q} \right ) \ \ .
\end{equation}
For $q$ real the quantity $n$ can be written as
\begin{equation}
n=\frac{1}{2 \log q} \log\left ( 
\frac{e^\epsilon - \, \sigma \, q^{-\sigma}}{e^\epsilon-\, \sigma \, 
q^{\sigma}} \right ) \ \ ,
\end{equation}
while for $q=e^{i \vartheta}$ complex as 
\begin{equation}
n=\frac{1}{\vartheta} \tan^{-1} \left ( 
\frac{\sin\vartheta}{e^\epsilon-\sigma\cos\vartheta} \right ) \ \ .
\end{equation}
We like to stress that to consider complex values of $q$ can be relevant 
in reproducing the exact interaction between particles in several physical 
systems. 
In Ref.\cite{sha2} it is analyzed the $q$-deformed pairing vibration 
in nuclei and is shown that the real part of $q$ simulates an attractive 
residual 
interaction between the nucleons and the imaginary part of $q$ decreases 
the binding energy of the pair of nucleons.

If we use as variable the single-particle energy $\epsilon$, it is easy 
to verify that 
the second order density fluctuation $\overline{(\Delta n)^2}=<n^2>-<n>^2$ 
can be expressed as \cite{ka2}
\begin{equation}
\overline{(\Delta n)^2}= \left \{ \frac{\partial}{\partial n} \log \left [ 
\frac{\varphi(n)}{\psi(n)}  \right] \right \}^{-1}\ \ ,
\end{equation}
or more explicitly for $q$-deformed bosons and fermions
\begin{equation}
\overline{(\Delta n)^2}=
\frac{\sinh\alpha}{\alpha} \, \, [n]\,[1+\, \sigma \, n] \ \ .
\end{equation}

The thermodynamic relations for $q$ oscillators can be derived starting from 
the explicit expression of the $\varphi(n)$ and $\psi(n)$ functions 
(see Ref.\cite{ka2} for details). The entropy density 
can be expressed as $S(n)=\int \log[\psi(n)/\varphi(n)] dn$ or more 
explicitly as 
\begin{equation}
S(n)=\int \log \frac{[1+\sigma n]}{[n]} \, dn \ \ .
\end{equation}

In conclusion, we have studied the kinetics of the $q$-oscillators defined in 
Eqs.(1)-(7) in the framework of a semiclassical non-linear approach. The two 
families of $q$-bosons and $q$-fermions can be interpreted as Brownian 
particles 
($J\propto v$, $D=const$). They obey to an exclusion-inclusion principle 
defined in 
terms of the transition probability by the functions  
$\varphi(n)=[n]$ and $\psi(n)=[1+\sigma n]$ which are implicitly 
related with the creation-annihilation operator matrix (see of Eqs.(10) 
and (11)). This relation fixes an important connection between quantum 
algebra and the semiclassical kinetics. \\
In this framework, we have derived a generalized Fokker Planck equation 
appropriate 
to study near-equilibrium and non-equilibrium $q$-oscillator systems. 
This equation 
is easily integrated in stationary condition and 
reproduces 
the statistical distributions for $q$-deformed bosons and fermions both for 
real and complex $q$ values.

\vspace{0.5cm}
\noindent
{\it We would like to thank M. Rego-Monteiro for stimulating discussions 
and for the critical reading of the manuscript.}\\

\vspace{0.5cm}
\noindent
{\it e-mails}: kaniadakis@polito.it, alavagno@polito.it and \\
quarati@polito.it

\end{document}